ACCEPTED FOR PUBLICATION IN MAGNETIC RESONANCE IN MEDICINE

# MR sequence design to account for non-ideal gradient performance

Daniel J West[1*], Felix Glang[2*], Jonathan Endres[3], David Leitão[1], Moritz Zaiss[3,4], Joseph V Hajnal[1,5], Shaihan J Malik[1,5]

[1]Imaging Physics & Engineering Research Department, School of Biomedical Engineering & Imaging Sciences, King's College London, London, United Kingdom
[2]Max-Planck-Institut fur biologische Kybernetik, Magnetic Resonance Center, Tübingen, Germany
[3]Institute of Neuroradiology, Erlangen University Hospital Institute of Neuroradiology, Erlangen, Germany
[4]Department of Artificial Intelligence in Biomedical Engineering, Friedrich-Alexander-Universität Erlangen-Nurnberg, Erlangen, Germany
[5]Early Life Imaging Research Department, School of Biomedical Engineering & Imaging Sciences, King's College London, London, United Kingdom

**Corresponding author**: Shaihan J Malik (shaihan.malik@kcl.ac.uk)

[*] *Daniel J West and Felix Glang contributed equally to this work*

## Abstract

*Purpose*: MRI systems are traditionally engineered to produce close to idealized performance, enabling a simplified pulse sequence design philosophy. An example of this is control of eddy currents produced by gradient fields; usually these are compensated by pre-emphasizing demanded waveforms. This process typically happens invisibly to the pulse sequence designer, allowing them to assume achieved gradient waveforms will be as desired. Whilst convenient, this requires system specifications exposed to the end-user to be substantially down-rated, since pre-emphasis adds an extra overhead to the waveforms. This strategy is undesirable for lower performance or resource-limited hardware. Instead, we propose an optimization-based method to design pre-compensated gradient waveforms that: (i) explicitly respect hardware constraints and (ii) improve imaging performance by correcting k-space samples directly.

*Methods*: Gradient waveforms are numerically optimized by including a model for system imperfections. This is investigated in simulation using an exponential eddy current model, then experimentally using an empirical gradient system transfer function on a 7T MRI system.

*Results*: Our proposed method discovers solutions that produce negligible reconstruction errors while satisfying gradient system limits, even when classic pre-emphasis produces infeasible results. Substantial reduction in ghosting artefacts from EPI imaging was observed, including an average reduction of 77% in ghost amplitude in phantoms.

*Conclusions*: This work demonstrates numerical optimization of gradient waveforms, yielding substantially improved image quality when given a model for system imperfections. While the method as implemented has limited flexibility, it could enable more efficient hardware usage, and may prove particularly important for maximizing performance of lower-cost systems.

## Introduction

MRI system hardware is traditionally engineered to produce close to *idealized performance*, enabling a simplified pulse sequence design philosophy in which the designer considers that the system will produce fields as demanded. A key requirement is that subsystems, particularly the radiofrequency (RF) transmit and gradient systems: (i) can operate independently from one time-period to the next, (ii) are precisely synchronized in time and (iii) respond linearly. A challenge to this ideal are gradient-induced eddy currents[1–5] (due to rapidly changing magnetic fields applied during MR acquisitions), gradient time delays and mechanical resonances, which can be characterized by a gradient system transfer function (GSTF).

To retain the *idealized performance* philosophy, these effects are often compensated by "pre-emphasizing" demanded waveforms after sequence definition but prior to amplification by gradient power amplifiers (GPAs). Though convenient, this necessitates enforcement of more stringent limits on hardware performance than the hardware can handle, since pre-emphasis adds an extra overhead to the waveforms. The *idealized performance* philosophy is effective, as evidenced by its universal adoption, but other methods that acknowledge and embrace true system performance are now feasible.

Assuming that an MR system is linear time-invariant, GSTFs are often applied during image reconstruction to perform post-hoc correction for gradient imperfections. Pre-acquisition correction strategies based on inversion of the GSTF have previously been demonstrated[6–8], where demanded gradient waveforms are pre-emphasized using frequency domain division by the GSTF. This is equivalent to more standard pre-emphasis methods but with the usual time domain decaying exponential model[9] replaced by a GSTF. Frequency domain division is, however, susceptible to zeros and GSTF inaccuracies, and cannot be done in a way that respects scanner hardware limits.

Both standard pre-emphasis and frequency domain correction lead to increased demanded slew rates, to counteract the smoothing effect of eddy currents. To respect hardware constraints, a typical solution 'down-rates' initial gradient waveforms, by designing them for a lower maximum slew rate and peak amplitude, such that once corrected, they remain within the system's actual capability. This approach was taken by Vannesjo *et al*.[6] for GSTF-based correction; whilst practical, the system’s resources are not optimally used. For scenarios with significant eddy currents, this results in substantial performance loss.

We propose an optimization-based method that designs pre-compensated gradient waveforms by considering a system distortion model, and explicitly respects hardware constraints, with a view to maximizing imaging performance. To gain more flexibility, rather than correcting time domain waveforms, we instead correct their zeroth-order moment (i.e. the k-space that they define) since this is ultimately the quantity of interest. The application of this method to echo planar imaging (EPI) sequences is explored using an exponential eddy current model and empirically measured GSTF (accounting for overall gradient performance).

## Theory

### Gradient system models and pre-emphasis

To model gradient imperfections, we distinguish between *demanded* control inputs and *realized* fields. We denote *demanded* waveforms as $\mathbf{g}(t)$, their temporal derivative as $\dot{\mathbf{g}}(t)$, and *realized* waveforms as $\mathbf{G}(t)$. To enable comparison with conventional sequence design strategies, both are expressed in the same units (mT/m), although $\mathbf{g}(t)$ would be realized as a current (or voltage) signal sent to GPAs. A common approach for quantifying (non-ideal) gradient performance is an exponential model[10]:

$$\mathbf{G}(t) = f\{\mathbf{g}(t)\} = \mathbf{g}(t) - \dot{\mathbf{g}}(t) \otimes \theta(t) \sum_n \alpha_n \exp\left(-\frac{t}{\tau_n}\right) \quad [1]$$

where $\alpha_n$ and $\tau_n$ are amplitude and time constants, $\theta(t)$ is a unit step function and $\otimes$ represents a convolution. To a first-order approximation, demanded waveforms can be pre-emphasized using[11]:

$$\mathbf{g}^{\mathrm{PRE}}(t) = \mathbf{g}(t) + \dot{\mathbf{g}}(t) \otimes \theta(t) \sum_n \tilde{\alpha}_n \exp\left(-\frac{t}{\tilde{\tau}_n}\right) \quad [2]$$

where $\tilde{\alpha}_n$ and $\tilde{\tau}_n$ are numerically optimized such that $f\{\mathbf{g}^{\mathrm{PRE}}(t)\} \approx \mathbf{g}(t)$ (i.e. $\tilde{\alpha}_n$ and $\tilde{\tau}_n$ deviate from true values of $\alpha_n$ and $\tau_n$ in the forward model).

Alternatively, gradient system response can be described using a GSTF, $\mathbf{H}(\omega)$ assuming linear time-invariance:

$$\widetilde{\mathbf{G}}(\omega) = \mathbf{H}(\omega)\tilde{\mathbf{g}}(\omega) \quad [3A]$$

where $\omega$ is temporal frequency and variables with tildes are frequency domain representations. This is written as a matrix to include cross-terms. The relationship can also be expressed in the time domain as a convolution:

$$\mathbf{G} = \mathbf{h}(t) \otimes \mathbf{g}(t) \quad [3B]$$

$$\begin{bmatrix} \mathrm{G}_x(t) \\ \mathrm{G}_y(t) \\ \mathrm{G}_z(t) \end{bmatrix} = \begin{bmatrix} h_{xx}(t) & h_{yx}(t) & h_{zx}(t) \\ h_{xy}(t) & h_{yy}(t) & h_{zy}(t) \\ h_{xz}(t) & h_{yz}(t) & h_{zz}(t) \end{bmatrix} \otimes \begin{bmatrix} \mathrm{g}_x(t) \\ \mathrm{g}_y(t) \\ \mathrm{g}_z(t) \end{bmatrix} \quad [3C]$$

where $\mathbf{h}(t)$ is the time domain representation of the GSTF (i.e. the gradient impulse response function, GIRF); cross-terms are written explicitly here. The GSTF can be measured empirically for an MR scanner *including* any pre-emphasis correction applied invisibly to the user. In principle, it may be extended to include both zero-order and higher-order terms[12], but these were not explored in this work for simplicity. Preliminary exploration indicated that first-order terms were the dominant source of error for sequences studied here. Additionally, the GSTF can be used for pre-emphasis by frequency domain division[7]:

$$\tilde{\mathbf{g}}^{\mathrm{PRE}}(\omega) = \mathbf{H}^{-1}(\omega)\tilde{\mathbf{g}}(\omega) \qquad [4]$$

<u>Optimization framework</u>

Pre-emphasis described by Equations 2 and 4 aims to produce realized gradients $\mathbf{G}(t)$ that closely match the desired, ideal behavior. Here, we instead optimize $\mathbf{g}(t)$ to produce k-space locations $\mathbf{k}_i \equiv \gamma \int_0^{t_\mathrm{i}} \mathbf{G}(t')\,dt'$ at sample times $t_\mathrm{i}$ that align with desired k-space sample locations $\mathbf{k_0}$. This is achieved by optimization:

$$\mathbf{g_{op}}(t) = \operatorname{argmin}\left\{\sum_{i=1}^{N_k}\left\|\mathbf{k_i} - \mathbf{k}_{\mathbf{0},\mathrm{i}}\right\|^2\right\} \qquad [5A]$$

subject to:

$$\left|\mathbf{g_{op}}(t)\right| \leq \mathrm{g}_{\max} \qquad [5B]$$

$$\left|\,\dot{\mathbf{g}}_{\mathbf{op}}(t)\right| \leq \mathrm{s}_{\max} \qquad [5C]$$

$\mathrm{g}_{\max}$ and $\mathrm{s}_{\max}$ are limits on demanded gradient amplitude and slew rate respectively. In practice, these constraints are realized by adding penalty terms to the optimization:

$$\mathbf{g_{op}}(t) = \operatorname{argmin}\left\{\sum_{i=1}^{\mathrm{N}_k}\left\|\mathbf{k}_\mathrm{i} - \mathbf{k}_{\mathbf{0},\mathrm{i}}\right\|^2 + w_\mathrm{g}\sum_{i=1}^{\mathrm{N}_t}\max(0, |\mathbf{g}(t_i)| - \mathrm{g}_{\max}) + w_{\dot{\mathrm{g}}}\sum_{i=1}^{\mathrm{N}_t}\max(0, |\dot{\mathbf{g}}(t_i)| - \mathrm{s}_{\max})\right\} \qquad [6]$$

where $\mathrm{N}_t$ is the number of timepoints, max operations enforce penalty terms that are only active when the constraints are violated and $w$ are tunable weighting factors.

## Methods

The proposed method was implemented as an extension to the MR-zero framework[13,14] and tested for several EPI sequence variants using simulated eddy currents (assuming an exponential model) and experimentally on a 7T MR scanner (Magnetom Terra, Siemens Healthineers, Forchheim, Germany) using a measured GSTF. The response function was measured using a modified thin-slice method with in-plane spatial encoding to capture cross-terms, as proposed by Rahmer *et al.*[15], using a spherical water phantom and a chirped test waveform.

Optimizations were run using the Adam[16] algorithm, taking advantage of pytorch/autograd to propagate derivatives through the loss function. Since the gradient distortion model (exponential or GSTF) is not a function of the gradient sample points being optimized, it can be treated as a constant from the perspective of propagation of derivatives. MR-zero with phase distribution graph[17] signal simulation was used to predict images obtained from pre- and post-optimization sequences with a 2D numerical brain phantom[18]. Optimized sequences were tested on both a spherical water phantom and healthy volunteer (female, age 25) who gave written consent under local ethical approval (HR-18/19-8700).

Example EPI sequences

Four variants of a single-slice 2D EPI sequence were explored: single-shot using acceleration factors R=1, 2 and 3 and multi-shot (16 shots, TR=5s, R=1). The sequence selection is arbitrary but use of these different acceleration factors and shot lengths changes the frequency content, making potential GSTF-mediated errors different. All sequences had a 250mm field-of-view; the single-shot R=1 sequence had matrix size 96; R=2 and multi-shot had matrix size 128; and R=3 had matrix size 129. The multi-shot sequence had TE=13ms. Partial Fourier sampling was used to maintain similar TEs for all single-shot scans (TE=33.8/33.6/36.6ms for R=1/2/3 cases respectively, with partial Fourier factors 73/96 for R=1 and 102/128 for R=2). An additional version of the R=1 sequence (termed shorter TE) was optimized using a shorter inter-echo spacing (1.05ms instead of 1.07ms); the nominal equivalent exceeds the scanner slew rate limit and so was not acquired.

Images were reconstructed using an inverse Fourier Transform (iFT) and basic GRAPPA implementation[19,20] where data were under-sampled. For these cases, a FLASH scan with matrix size 128x128 was used to obtain calibration data, with GRAPPA kernel size [kx,ky]=[9,5]. Partial Fourier-induced Gibbs ringing was reduced using a smooth-step filter. No phase corrections were made for ghost correction of EPI data.

Simulated experiments with exponential eddy current model

Single-shot (R=1) EPI optimizations assumed a scanner-typical eddy current term ($\tau$=50μs). Its scaling parameter was adjusted to produce a significant but not unrealistic perturbation ($\alpha$=5e-6). Optimization of the 17,040 gradient samples was performed on a 20(40)×Intel(R) Xeon(R) Silver 4210 2.20GHz CPU, 251GB RAM, 32GB NVIDIA Tesla V100 GPU (NVIDIA, Santa Clara, CA, USA), with realistic scanner hardware limits and a learning rate of 5e-5. The optimization result was compared to that obtained using pre-emphasis (with separately optimized eddy current properties: $\tilde{\alpha}$=1e-5 and $\tilde{\tau}$=22.2μs) using normalized root-mean-square error (NRMSE) of pixel intensities calculated with respect to the target image.

Physical experiments using MR system with empirical GSTF

Optimizations of all four EPI variants were completed using Equation 3B as the hardware model (only first-order terms were considered). Plots of measured terms for different physical axes are in Supporting Information Figure S1. Scanner amplitude and slew rate limits of $g_{max}$=72mT/m and $s_{max}$=180mT/m/ms were enforced during optimization. Optimization solutions are compared to those obtained when inverting the GSTF (Equation 4) using NRMSE. GSTF-based pre-emphasis was computed using only self-terms and a Tukey filter (45kHz cut-off frequency, 0.3 transition width) to suppress noise amplification due to zeroes at high GSTF frequencies.

Since the GSTF was measured using the scanner's own pre-emphasis correction, it characterizes the system response when operating under normal conditions. All EPI experiments were run under these conditions, hence both GSTF-based pre-emphasis and optimized solutions should be understood as *additional* corrections, beyond any existing gradient corrections the scanner uses.

EPI gradient waveforms were discretized on a time grid matching the gradient raster time (10μs). A learning rate of 5e-6 was used with 50,000 iterations for all single-shot cases (taking ~40 mins with ~0.05s per iteration) and 20,000 iterations for multi-shot cases. All sequences were implemented using Pulseq[21,22]. Ghosting artefacts were quantified by taking the difference between RMS signal outside the object in phase-encode and frequency-encode directions, normalized to RMS signal in the phantom.

# Results

Simulations using exponential eddy current model

Figure 1 shows optimized gradient waveforms compared to the standard pre-emphasis solution in simulation. Both show similar characteristics, but the optimized solution does not violate the defined slew rate limit (180mT/m/ms). Sample locations are restored to a regular Cartesian grid using either approach, as shown in the plotted k-space portions, with negligible image reconstruction error (<0.2%) for both.

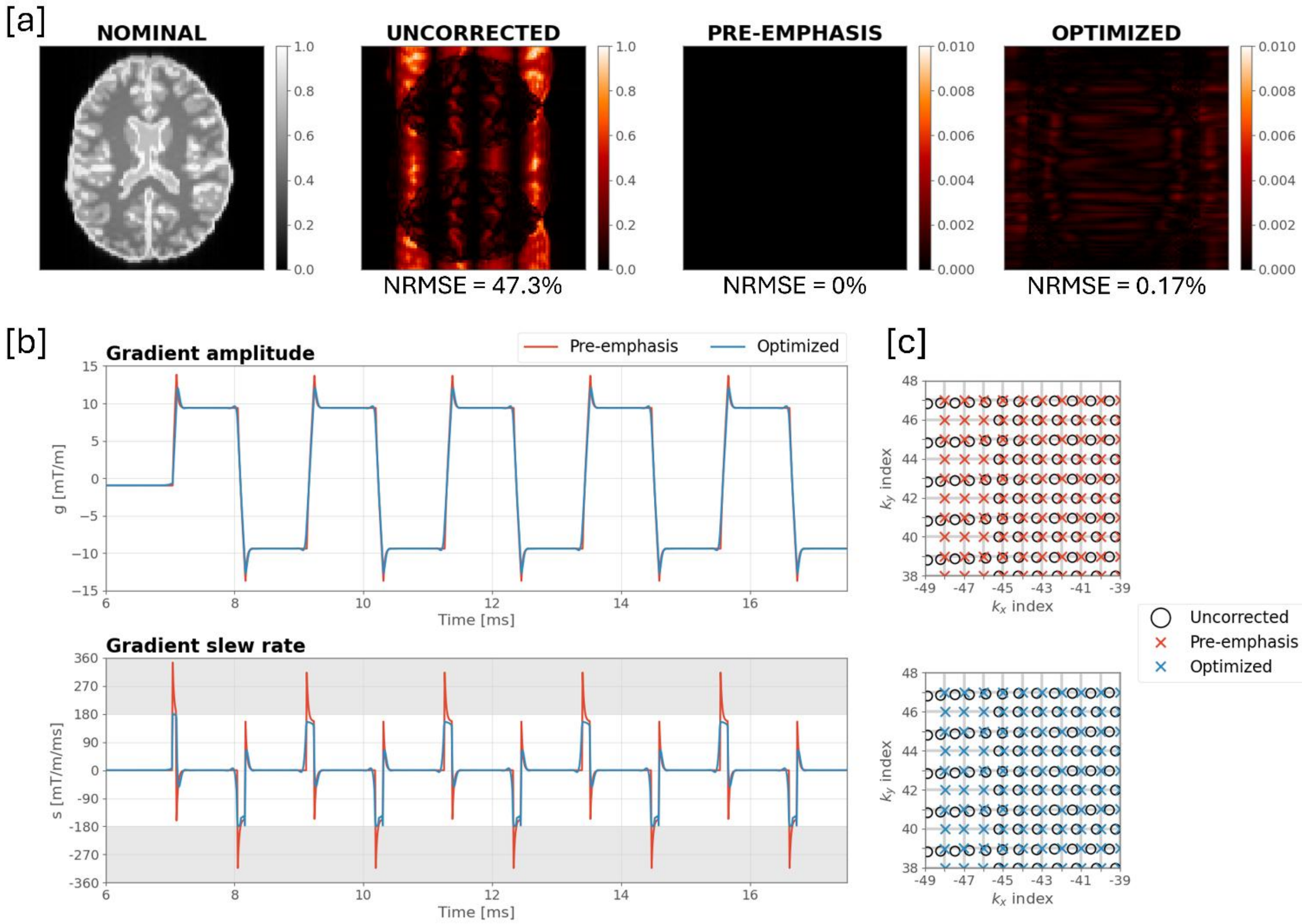

Figure 1: Comparison of our optimization results to standard pre-emphasis when using a simulated exponential eddy current model with τ=50μs. The nominal image reconstruction (no gradient errors) is shown in [a] alongside corresponding differences when eddy currents are included before correction, after pre-emphasis, and after optimization (NRMSEs displayed underneath). [b] Demanded gradients and corresponding slew rates are plotted over a zoomed timescale to highlight waveform features, with k-space locations shown in [c] (gray lines correspond to the desired k-space grid). Optimization finds a solution with almost perfectly restored k-space samples whilst satisfying scanner hardware constraints. Pre-emphasis also corrects the k-space locations but exceeds slew rate constraints (forbidden regions are shaded gray in gradient slew rate plot).

Simulations using GSTF model

Figure 2 illustrates k-space displacements caused by gradient imperfections as modelled by the GSTF for examples of single-shot and multi-shot EPI sequences. Odd and even lines/shots are not just offset from one another; different degrees of displacement are seen throughout k-space. The proposed optimization and GSTF-based pre-emphasis aim to eliminate these shifts.

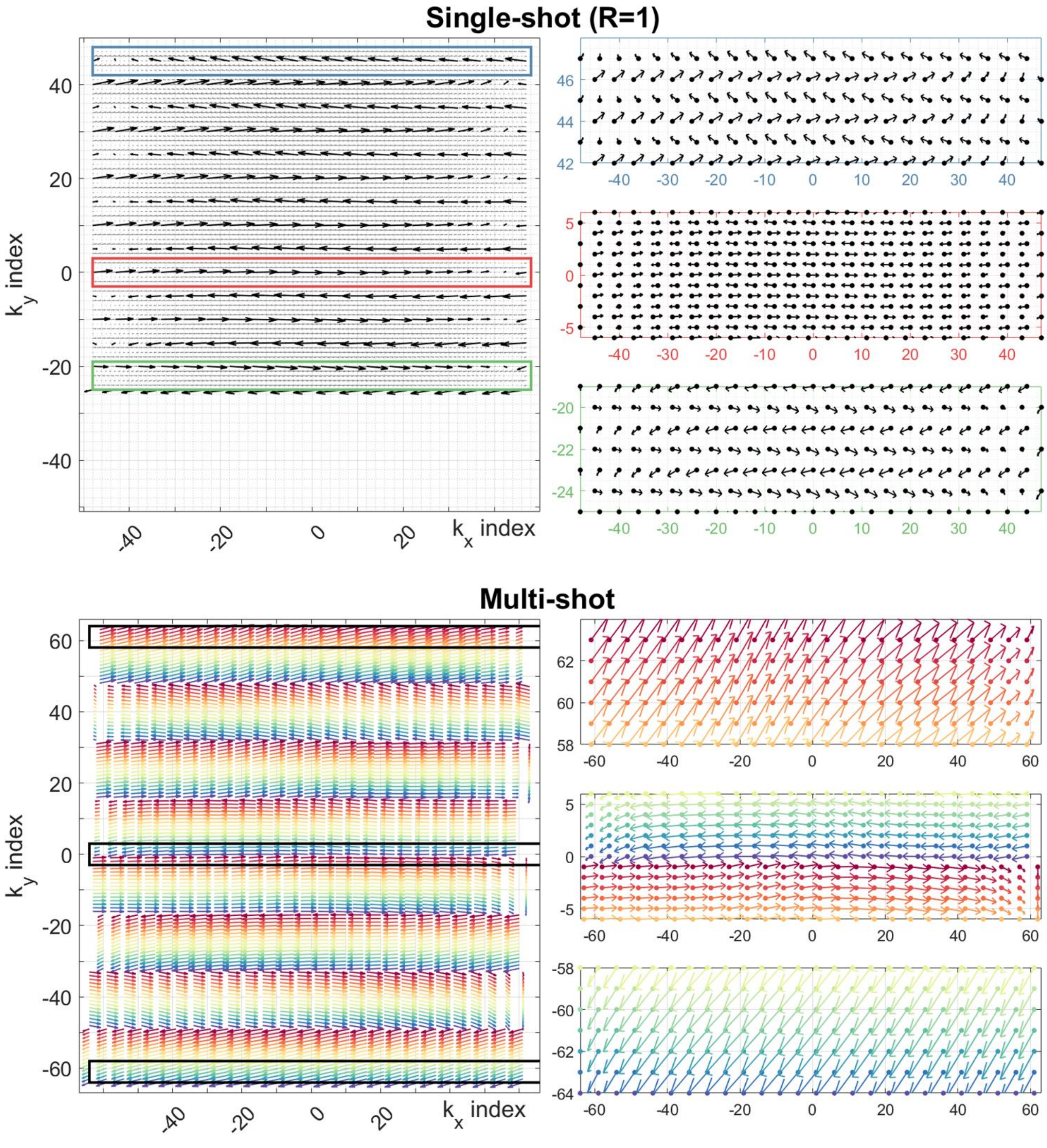

Figure 2: k-space displacements indicated by arrows pointing from nominal demanded locations to realized locations for entire k-spaces (LHS) and zoomed k-space regions to visualize displacement more clearly, with dots indicating nominal demanded locations (RHS). As arrow sizes and directions are spatially inconsistent, post-hoc phase shifts would be unsuccessful in mitigating image imperfections. Down-sampling of arrows is performed for clarity. Multi-shot arrows are colored according to shot number. Partial Fourier is visible in the top LHS for the single-shot example.

Figure 3 illustrates simulated images for the single-shot (R=1) EPI sequence. When no correction is applied, images show significant ghosting artefacts. GSTF-based pre-emphasis yields a solution with low error (0.39%) that exceeds the slew rate limit. Optimization achieves a solution with lower error than pre-emphasis (0.11%) whilst respecting gradient system constraints. This is achieved using subtle waveform changes that do not resemble the more dramatic 'over-driven' pre-emphasis solution (red line in Figure 3d). Note that pre-emphasis using the simple exponential eddy current model (Figure 1a) returns a slightly lower error than our optimization whereas in Figure 3, pre-emphasis produces slightly higher error. Given that pre-emphasis is not subject to any constraints, it might be expected that this would achieve lower error; for the empirical GSTF this is not the case because of filtering used to limit noise from high frequencies in the GSTF. In both examples, the proposed optimization returns low error while also satisfying constraints.

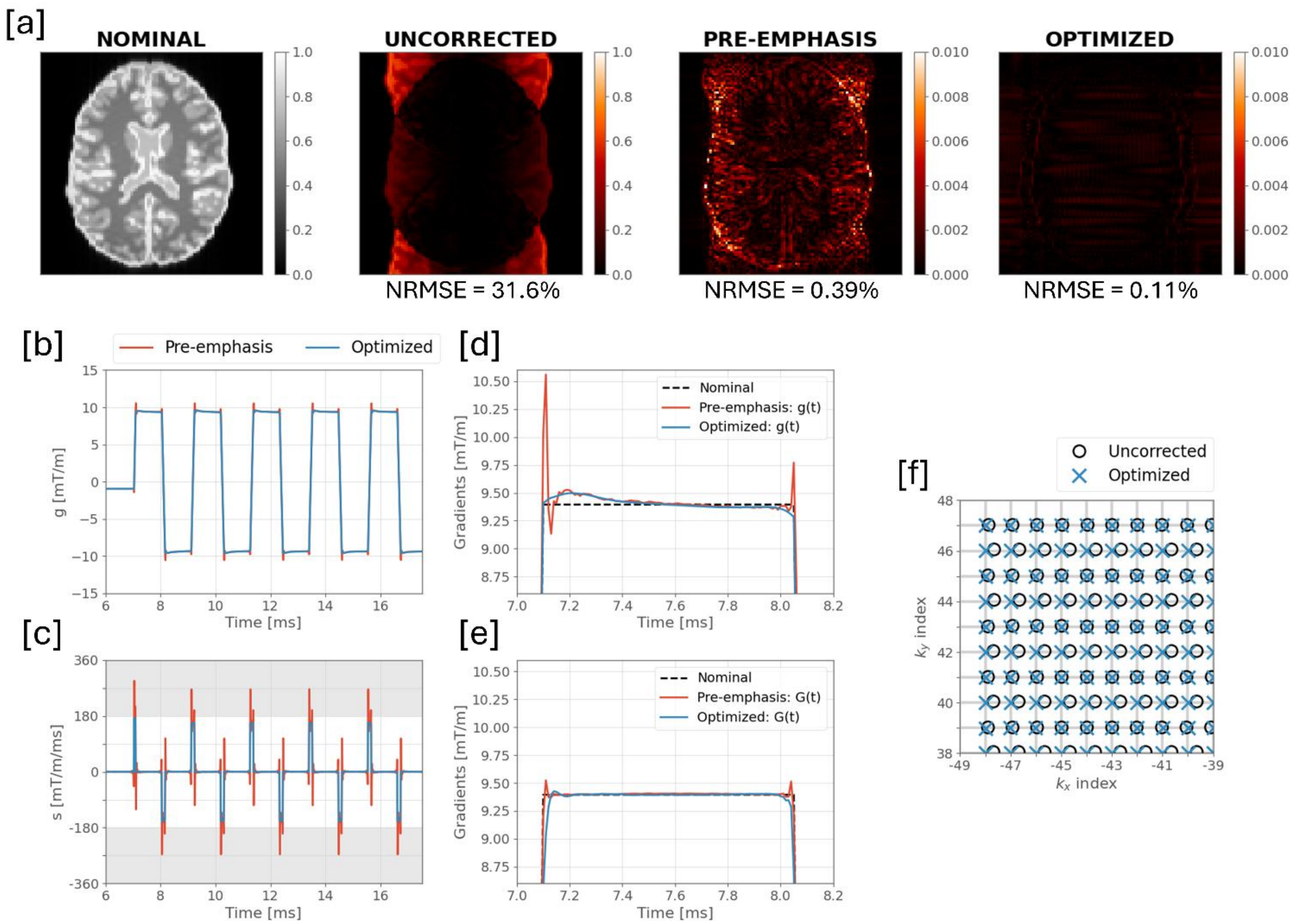

Figure 3: [a] Simulated images for a numerical brain phantom using the 7T GSTF and single-shot (R=1) EPI sequence. The nominal (target) image does not consider gradient imperfections and the uncorrected image shows error predicted from GSTF without additional correction. GSTF-based pre-emphasis yields low error but exceeds the slew rate limit (180mT/m/ms) [c]. The right-hand pane in [a] shows equivalent error from optimization and its solution (blue demanded gradient traces in [b]) does not violate slew rate limits [c]. Corresponding NRMSEs are shown underneath

each difference map. Zoomed optimized and GSTF-based pre-emphasis demanded [d] and realized [e] waveforms for one readout. The optimization solution remains within system limits and corrects k-space locations [f].

Experimental results

Figure 4 depicts solutions for all EPI sequence variants described above; part (a) displays simulated images using the numerical brain phantom (also used in Figures 1 and 3), while part (b) shows data acquired from the 7T system using a spherical water phantom. The phantom shape is slightly distorted due to local off-resonance caused by the phantom holder, and signal shading is due to the receiver bias field which has been left uncorrected; neither is relevant to this discussion. Simulations (Figure 4a) predict significant ghosting in uncorrected image reconstructions for all variants that are eliminated by the optimization. The acquired phantom images show similar results, with uncorrected sequences yielding similar artefacts that are substantially reduced for the optimized sequences (indicated by decreases in relative artefact intensity in Figure 4d). Figure 4c shows that peak demanded slew rate always exceeds the limit ($s/s_{\mathrm{max}}$=1) using pre-emphasis, so these solutions were not tested experimentally. Our method satisfies constraints for all scenarios, including the shorter TE sequence which could run using a shorter inter-echo spacing than possible with the nominal sequence.

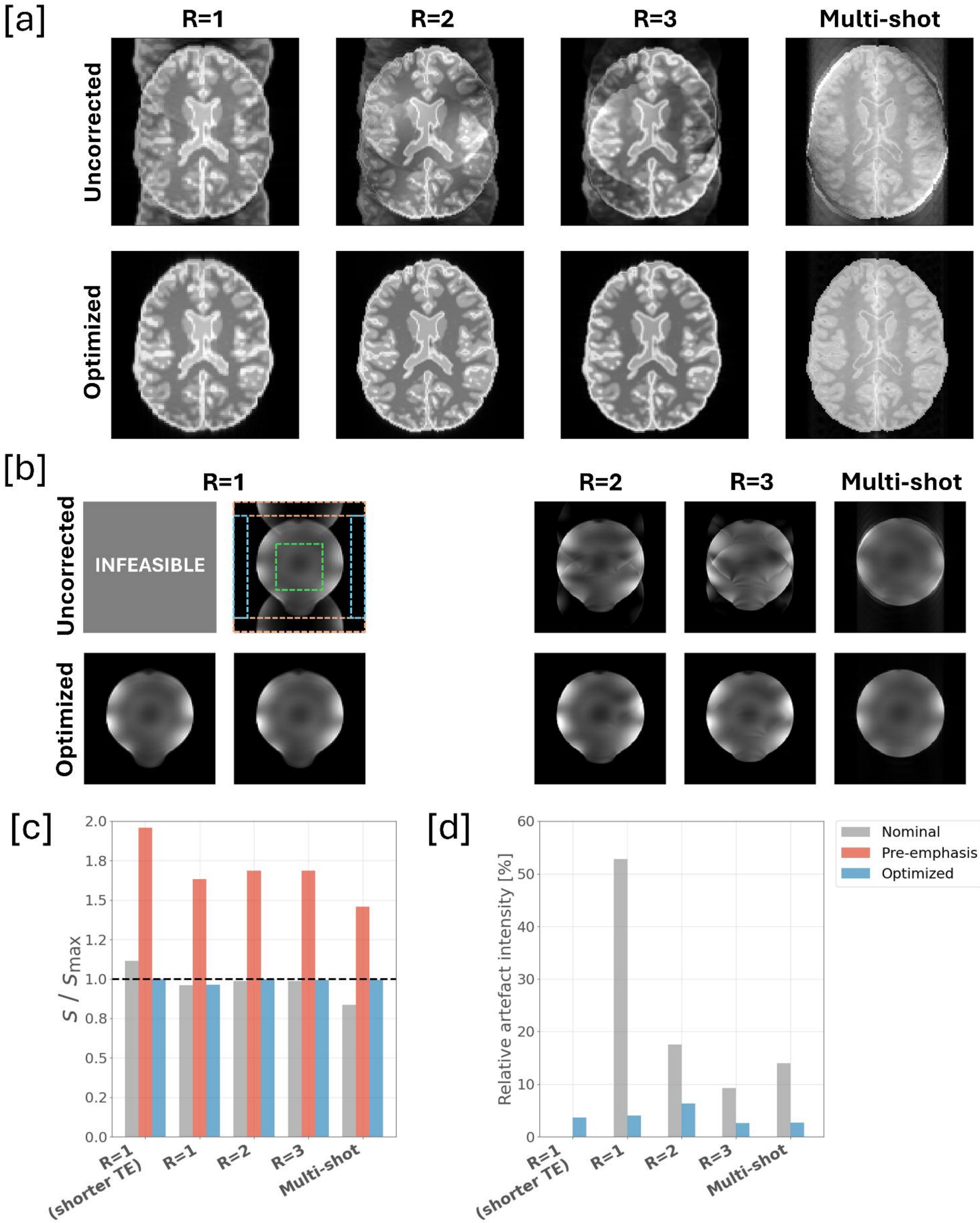

Figure 4: [a] Simulated and [b] experimental phantom results for EPI variants: R=1,2,3 and multi-shot using the 7T system. Simulated and experimentally acquired uncorrected sequences show image artefacts that are very similar in appearance. Optimization substantially reduces artefacts both in simulation and experiment. Experimental data contains two variants of the fully sampled (R=1) case, one with shorter TE (LHS); this violates the slew rate limit, so an uncorrected image was not acquired, whereas optimization finds a solution that delivers this lower TE without ghosting. [c] Peak slew rate for each sequence, showing that our method always meets the specified constraint and [d] relative artefact intensities computed from the outlined regions in [b] (difference between RMS signal in orange and blue boxes, normalized to RMS signal in green box). Pre-emphasis using GSTF violates the slew rate limit for all cases, so was not tested experimentally.

The same sequences were tested in vivo with results presented in Figure 5 using different window settings to more clearly reveal the ghosts in uncorrected images. Ghosting is again substantially reduced for all variants, most notably for R=1. Artefacts manifest themselves differently for multi-shot EPI but are also suppressed after optimization.

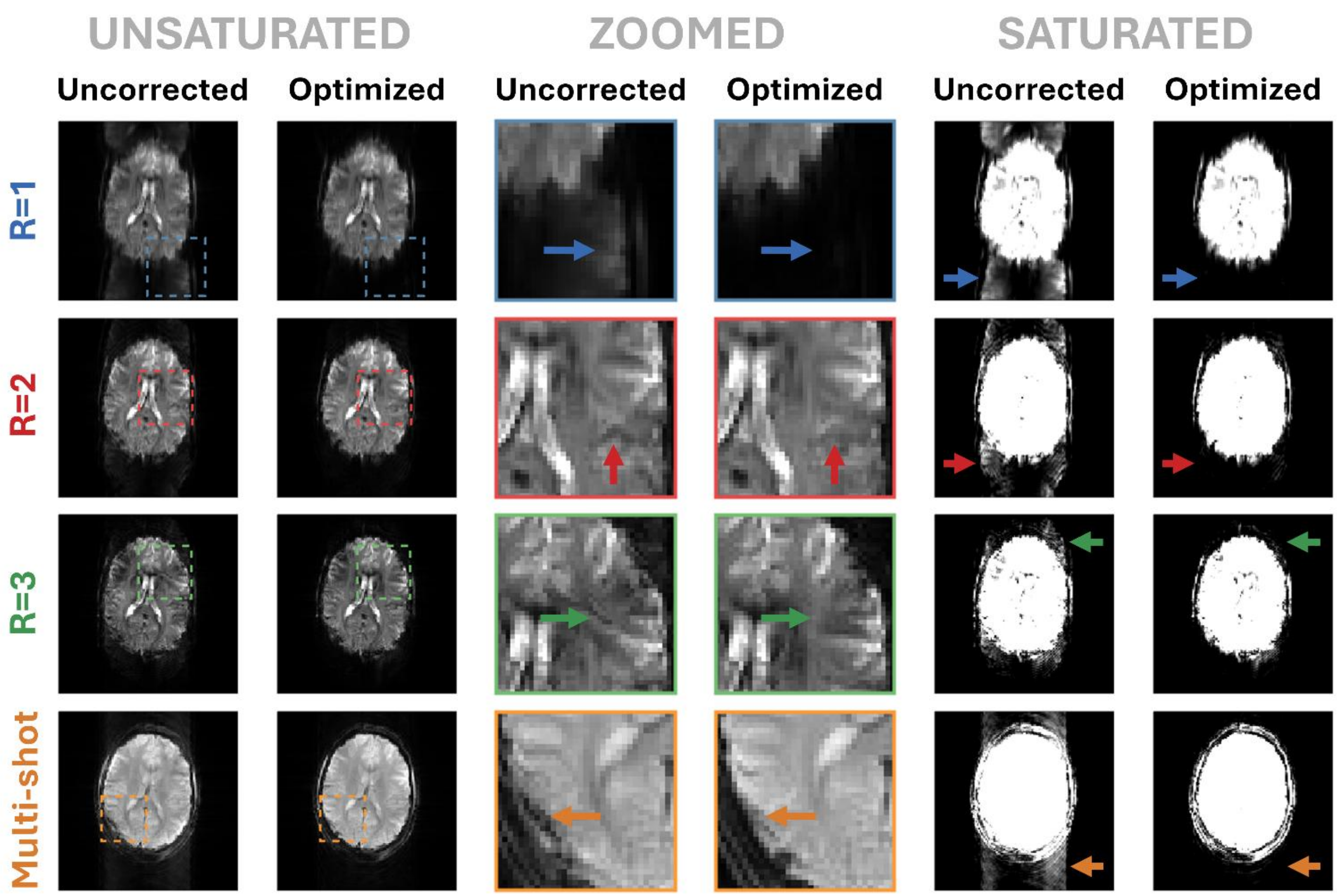

Figure 5: In vivo results for all EPI variants. Ghosting artefacts inside and outside the brain are removed following optimization, as demonstrated by zoomed portions for each color-coded EPI variant (middle two columns) and full-image subplots with saturated color ranges (two right-most columns). Color-coded arrows indicate regions where the improvement with the proposed method is obvious. Identical color ranges are used for images acquired using corresponding uncorrected and optimized sequences.

## Discussion and Conclusions

This work proposes an optimization-based method to design gradient waveforms, compensating for system imperfections via constrained optimization including an empirical model for gradient system performance. The approach differs from pre-emphasis by: (i) allowing amplitude/slew rate constraints to be respected, and (ii) considering the achieved k-space sampling, not the gradient waveform itself, during optimization. We explored use of a simple exponential eddy current model and a GSTF as the forward model, presenting both simulated images and experimental data that demonstrate image quality improvements using the proposed method. Since our approach requires optimization of sequences prior to scanning, it is unsuitable for existing system architectures where user-defined resolution/fields-of-view dynamically change demanded gradients, requiring on-the-fly correction. Rather, this work is a proof-of-concept, where we explore the outlook and potential use cases below.

Figure 1 demonstrates an example for a short-term ($\tau$=50μs) eddy current, where classic pre-emphasis (Equation 2) produces a large slew rate spike that cannot be achieved in practice. We demonstrate that numerical optimization can achieve correct k-space sampling while respecting hardware limits. Moving to a real-world example, Figures 2-3 focus on using the GSTF of a 7T system *including* all vendor-implemented gradient corrections. Figure 2 depicts realized k-space sampling from uncorrected sequences; errors are more complex than simple delays between adjacent lines. GSTF-based pre-emphasis[6–8] can correct these deviations using frequency domain division, but as with the simpler eddy current model, resulting waveforms are likely to exceed physical constraints. The proposed method corrects trajectory-related errors by directly optimizing gradient waveforms based on the achieved k-space, rather than the distorted gradient waveforms themselves. The solutions found for several EPI sequences do not resemble the 'over-driven' appearance of pre-emphasis (e.g. Figures 3b-e), instead featuring more subtle modulations. The ability to work within constraints is a key benefit. Others have noted this limitation of pre-emphasis, proposing to either 'down-rate' the initial waveforms[6] (which compromises performance and is not optimal) or use iterative calculations with a time-stretching algorithm to enforce physical constraints[23].

This work is an acquisition-based correction for imperfections. Reconstruction-based, retrospective correction is also possible; indeed, simple phase correction of odd and even lines in an EPI acquisition is an example of this, as are more sophisticated methods using a GSTF[24–27] or temporal convolutional networks[28]. In this proof-of-concept study, we only use iFT reconstruction (excluding any post-acquisition correction methods), to focus on sampling fidelity. Ultimately, we envisage prospective and retrospective corrections working together. GSTF-based optimization requires system stability to be effective. The success of our experimental work is testament to the fact that gradient systems in high performance scanners tend to be very stable, as noted by others who use GSTFs in image reconstruction[24–27]. There are data showing temperature-dependence of the gradient response, but this effect can also be modelled[29]. Although our optimizations were time-consuming, the same solutions can be used in multiple experiments; phantom and in vivo data were obtained using identical gradient waveforms optimized with a GSTF acquired days earlier.

Others have also proposed gradient correction methods that are constrained by hardware limits. Harkins *et al.* explored pre-distortion of gradient waveforms to correct gradient errors in RF pulses[30] and proposed an iterative correction scheme that includes repeated measurement of waveforms (useful when systems are less stable or not linear time-invariant). The GrOpt toolbox[31,32] enables gradient waveform design under amplitude and slew rate constraints (among others). Applications have included individual preparation modules (e.g. diffusion- or velocity-encoding gradients) that consider different properties of the gradient waveforms in the objective function. For diffusion, b-value is the objective, whilst for phase contrast MRI, background phase error (due to the zeroth-order moment) is the objective[33]. Our work is related, since the k-space objective function is also focused on the zeroth-order moment of the designed waveforms, but we instead focus on whole EPI readouts rather than individual preparation modules.

The proposed optimization-based approach must be performed for a specific field-of-view (including angulation, and for a fixed resolution). Modifying the waveforms retrospectively, using rotation matrices to rotate the field-of-view for example, is not guaranteed to produce optimal results. This is very different from the current paradigm in MRI that uses corrections applied on-the-fly to enable maximum user-flexibility. Future work will explore parameterized solutions that can be adapted to enable some user-flexibility (e.g. field-of-view modification). Alternatively, future scanners could use fixed geometries to obtain acceptable performance from lower-cost hardware.

We used the Adam optimizer, which is more typically used to train neural networks, since it was found empirically to work well (it is a gradient-based method that performs well in high-dimensional, non-convex problems). Other optimizers will be explored in the future, as will a strategy that takes advantage of the time-sequential nature of gradient waveforms. In terms of hardware models, we considered an analytic eddy current model, alongside an empirical GSTF that describes the response of the whole chain including GPAs and the controlling logic implemented on them (e.g. feedback control). We aim to develop more sophisticated models that include characteristics of the gradient power supply (e.g. voltage and charge capacity, and thermal models). This approach may potentially allow hardware to be driven closer to its true limits, as will inclusion of other effects such as acoustic resonances and peripheral nerve stimulation.

## Acknowledgements

This work was supported by core funding from the Wellcome/EPSRC Centre for Medical Engineering [WT203148/Z/16/Z], by a Wellcome Trust Collaboration in Science Award [WT 201526/Z/16/Z] and by the National Institute for Health Research (NIHR) Clinical Research Facility based at Guy's and St Thomas' NHS Foundation Trust and King's College London. We are grateful for the support of the Max Planck Society. The views expressed are those of the author(s) and not necessarily those of the NHS, the NIHR or the Department of Health and Social Care.

## Data Availability Statement

The source code used for simulation studies and to generate sequences for in silico and in vivo experiments is available from: https://github.com/mriphysics/Optimization-for-non-ideal-gradients.

## References

1. Ahn CB, Cho ZH. Analysis of the Eddy-Current Induced Artifacts and the Temporal Compensation in Nuclear Magnetic Resonance Imaging. *IEEE Trans Med Imaging*. 1991;10(1):47-52. doi:10.1109/42.75610

2. Doty FD. MRI Gradient Coil Optimization. In: Blumler P, Blumich B, Botto R, Fukushima E, eds. *Spatially Resolved Magnetic Resonance Methods, Materials, Medicine, Biology, Rheology, Geology, Ecology, Hardware*. Wiley; 1998:647-674.

3. Schmitt F. The Gradient System. *Proc Intl Soc Mag Reson Med 21*. 2013:1-13.

4. Spees WM, Buhl N, Sun P, Ackerman JJH, Neil JJ, Garbow JR. Quantification and Compensation of Eddy-Current-Induced Magnetic Field Gradients. *Journal of Magnetic Resonance*. 2011;212(1):116-123. doi:10.1016/j.jmr.2011.06.016

5. Trakic A, Liu F, Sanchez Lopez H, Wang H, Crozier S. Longitudinal gradient coil optimization in the presence of transient eddy currents. *Magn Reson Med*. 2007;57(6):1119-1130. doi:10.1002/mrm.21243

6. Vannesjo SJ, Duerst Y, Vionnet L, et al. Gradient and shim pre-emphasis by inversion of a linear time-invariant system model. *Magn Reson Med*. 2017;78(4):1607-1622. doi:10.1002/mrm.26531

7. Stich M, Wech T, Slawig A, et al. Gradient waveform pre-emphasis based on the gradient system transfer function. *Magn Reson Med*. 2018;80(4):1521-1532. doi:10.1002/mrm.27147

8. Scholten H, Lohr D, Wech T, Köstler H. Fast measurement of the gradient system transfer function at 7 T. *Magn Reson Med*. 2023;89(4):1644-1659. doi:10.1002/mrm.29523

9. Bernstein MA, King KF, Xiaohong JZ. *Handbook of MRI Pulse Sequences*. Academic Press; 2004. doi:https://doi.org/10.1016/B978-0-12-092861-3.X5000-6

10. Jehenson P, Westphal M, Schuff N. Analytical method for the compensation of eddy-current effects induced by pulsed magnetic field gradients in NMR systems. *Journal of Magnetic Resonance (1969)*. 1990;90(2):264-278. doi:10.1016/0022-2364(90)90133-T

11. Bernstein MA, King KF, Xiaohong JZ. *Handbook of MRI Pulse Sequences*. Academic Press; 2004. doi:https://doi.org/10.1016/B978-0-12-092861-3.X5000-6

12. Vannesjo SJ, Haeberlin M, Kasper L, et al. Gradient system characterization by impulse response measurements with a dynamic field camera. *Magn Reson Med*. 2013;69(2):583-593. doi:10.1002/mrm.24263

13. Loktyushin A, Herz K, Dang N, et al. MRzero - Automated discovery of MRI sequences using supervised learning. *Magn Reson Med*. 2021;86(2):709-724. doi:10.1002/mrm.28727

14. Glang F, Loktyushin A, Herz K, et al. Advances in MRzero - supervised learning of parallel imaging sequences including joint non-Cartesian trajectory and flip angle optimization. In: *Proc. Intl. Soc. Mag. Reson. Med. 29*. ; 2021.

15. Rahmer J, Mazurkewitz P, Börnert P, Nielsen T. Rapid acquisition of the 3D MRI gradient impulse response function using a simple phantom measurement. *Magn Reson Med*. 2019;82(6):2146-2159. doi:10.1002/mrm.27902

16. Kingma DP, Ba JL. Adam: a method for stochastic optimization. *ArXiv*. 2014;1412.6980:1-15.

17. Endres J, Weinmüller S, Dang HN, Zaiss M. Phase distribution graphs for fast, differentiable, and spatially encoded Bloch simulations of arbitrary MRI sequences. *Magn Reson Med*. 2024. doi:10.1002/mrm.30055

18. Cocosco CA, Kollokian V, Kwan RKS, Pike GB, Evans AC. BrainWeb:Online Interface to a 3D MRI Simulated Brain Database. *Neuroimage*. 1997;5(4). http://www.bic.mni.mcgill.ca/users/crisco/HBM97_abs/HBM97_abs.pdf%0Ahttp://www.bic.mni.mcgill.ca/users/crisco/HBM97_poster/HBM97_poster.pdf.

19. Griswold MA, Jakob PM, Heidemann RM, et al. Generalized Autocalibrating Partially Parallel Acquisitions (GRAPPA). *Magn Reson Med*. 2002;47(6):1202-1210. doi:10.1002/mrm.10171

20. Breuer FA, Kannengiesser SAR, Blaimer M, Seiberlich N, Jakob PM, Griswold MA. General formulation for quantitative G-factor calculation in GRAPPA reconstructions. *Magn Reson Med*. 2009;62(3):739-746. doi:10.1002/mrm.22066

21. Layton KJ, Kroboth S, Jia F, et al. Pulseq: A rapid and hardware-independent pulse sequence prototyping framework. *Magn Reson Med*. 2017;77(4):1544-1552. doi:10.1002/mrm.26235

22. Ravi K, Geethanath S, Vaughan J. PyPulseq: A Python Package for MRI Pulse Sequence Design. *J Open Source Softw*. 2019;4(42):1725. doi:10.21105/joss.01725

23. Çavuşoğlu M, Mooiweer R, Pruessmann KP, Malik SJ. VERSE-guided parallel RF excitations using dynamic field correction. *NMR Biomed*. 2017;30(6). doi:10.1002/nbm.3697

24. Campbell-Washburn AE, Xue H, Lederman RJ, Faranesh AZ, Hansen MS. Real-time distortion correction of spiral and echo planar images using the gradient system impulse response function. *Magn Reson Med*. 2016;75(6):2278-2285. doi:10.1002/mrm.25788

25. Lee NG, Ramasawmy R, Lim Y, Campbell-Washburn AE, Nayak KS. MaxGIRF: Image reconstruction incorporating concomitant field and gradient impulse response function effects. *Magn Reson Med*. 2022;88(2):691-710. doi:10.1002/mrm.29232

26. Kronthaler S, Rahmer J, Börnert P, et al. Trajectory correction based on the gradient impulse response function improves high-resolution UTE imaging of the musculoskeletal system. *Magn Reson Med*. 2021;85(4):2001-2015. doi:10.1002/mrm.28566

27. Vannesjo SJ, Graedel NN, Kasper L, et al. Image reconstruction using a gradient impulse response model for trajectory prediction. *Magn Reson Med*. 2016;76(1):45-58. doi:10.1002/mrm.25841

28. Martin JB, Alderson HE, Gore JC, Does MD, Harkins KD. Modeling the MRI gradient system with a temporal convolutional network: Improved reconstruction by prediction of readout gradient errors. *Magn Reson Med*. 2025. doi:10.1002/mrm.70044

29. Nussbaum J, Dietrich BE, Wilm BJ, Pruessmann KP. Thermal variation in gradient response: measurement and modeling. *Magn Reson Med*. 2022;87(5):2224-2238. doi:10.1002/mrm.29123

30. Harkins KD, Does MD, Grissom WA. Iterative method for predistortion of MRI gradient waveforms. *IEEE Trans Med Imaging*. 2014;33(8):1641-1647. doi:10.1109/TMI.2014.2320987

31. Loecher M, Middione MJ, Ennis DB. A gradient optimization toolbox for general purpose time-optimal MRI gradient waveform design. *Magn Reson Med*. 2020;84(6):3234-3245. doi:10.1002/mrm.28384

32. Middione MJ, Loecher M, Moulin K, Ennis DB. Optimization methods for magnetic resonance imaging gradient waveform design. *NMR Biomed*. 2020;33(12). doi:10.1002/nbm.4308

33. Loecher M, Ennis DB. Phase contrast MRI with minimized background phase errors. *Magn Reson Med*. 2025;93(3):1104-1116. doi:10.1002/mrm.30336

# Supporting Information

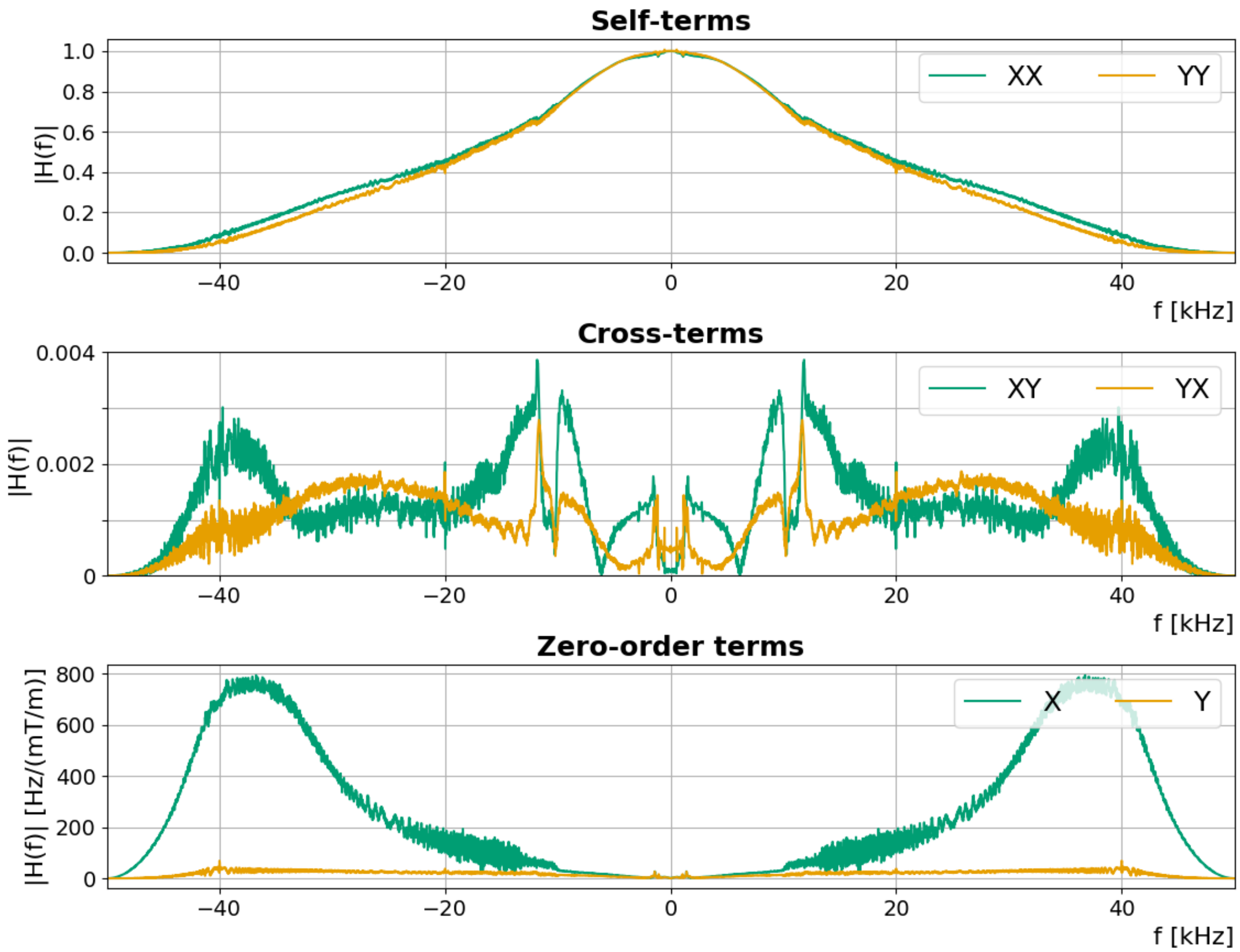

S1: GSTF first-order self-terms (top) and cross-terms (middle). The latter are far smaller (note the change in y-axis scale), however the system model used for optimization includes all four terms for completeness. Zero-order terms are displayed in the bottom sub-plot for reference but were not used in this work since their impact on the sequences studied was found to be small.